\documentclass[12pt]{article} 
\usepackage{cite} 

\usepackage{latexsym}
\usepackage{amsmath}
\usepackage{amssymb}
\usepackage{pstricks}
\usepackage{indentfirst}
\usepackage{amsthm}
\theoremstyle{definition}

\newcommand{\ad}{^\dagger }

\newcommand{\becs}{\begin{cases}}
\newcommand{\bem}{\begin{matrix}}

\newcommand{\dya}[1]{|#1\rgl\lgl#1|}
\newcommand{\dyad}[2]{|#1\rgl\lgl#2|}

\newcommand{\encs}{\end{cases}}
\newcommand{\enm}{\end{matrix}}

\newcommand{\hf}{{\textstyle\frac{1}{2} }}

\newcommand{\inpd}[2]{\lgl#1|#2\rgl }
\newcommand{\ket}[1]{|#1\rgl }
\newcommand{\lbrk}[1]{\left\{\vrule height #1cm depth #1cm width 0pt\right.}

\newcommand{\lgl}{\langle } 
\newcommand{\lp}{\left(} 

\newcommand{\mte}[2]{\lgl#1|#2|#1\rgl }
\newcommand{\mted}[3]{\lgl#1|#2|#3\rgl }

\newcommand{\ot}{\otimes }

\newcommand{\rbrk}[1]{\left.\vrule height #1cm depth #1cm width 0pt\right\}}
\newcommand{\rgl}{\rangle }

\newcommand{\rp}{\right)}

\newcommand{\st}{\sqrt{2}}

\newcommand{\Tr}{{\rm Tr}}

\newcommand{\CS}{{\cal C}}

\newcommand{\GS}{{\cal G}}
\newcommand{\HS}{{\cal H}}

\newcommand{\PS}{{\cal P}}

\newcommand{\ST}{{\cal S}}

\newcommand{\VS}{{\cal V}}
\newcommand{\WS}{{\cal W}}
\newcommand{\XS}{{\cal X}}
\newcommand{\YS}{{\cal Y}}
\newcommand{\ZS}{{\cal Z}}

\newcommand{\al}{\alpha }
\newcommand{\bt}{\beta }
\newcommand{\gm}{\gamma }
\newcommand{\Gm}{\Gamma }
\newcommand{\dl}{\delta }

\newcommand{\Thm}{\textbf{Theorem\ }}
\newcommand{\Lem}{\textbf{Lemma\ }}

\begin{document}

\title{Types of Quantum Information}

\author{Robert B. Griffiths\\
Physics Department,
Carnegie-Mellon University,\\
Pittsburgh, PA 15213, USA}

\date{Version of 10 September 2007}

\maketitle  

\begin{abstract}
  Quantum, in contrast to classical, information theory, allows for different
  incompatible types (or species) of information which cannot be combined with
  each other.  Distinguishing these incompatible types is useful in
  understanding the role of the two classical bits in teleportation (or one
  bit in one-bit teleportation), for discussing decoherence in
  information-theoretic terms, and for giving a proper definition, in quantum
  terms, of ``classical information.'' Various examples (some updating earlier
  work) are given of theorems which relate different incompatible kinds of
  information, and thus have no counterparts in classical information theory.
\end{abstract}


	\section{Introduction}
\label{sct1}

Despite an enormous number of publications in the field of quantum information
(see \cite{NlCh00,Kyl02} for useful introductions), neither the fundamental
principles underlying the subject, nor its connection with classical
information theory as developed by Shannon and his successors \cite{CvTh91},
is altogether clear.  On the one hand there has been some dispute
\cite{BrZl01,Tmps03} about whether Shannon's ideas can be applied at all in
the quantum domain.  On the other hand there have been suggestions that the
connection with Shannon's ideas occurs only for macroscopic systems or
asymptotically large $N$ (number of transmissions, or whatever) limits, as in
what is sometimes called ``quantum Shannon'' theory \cite{DvHW05,DvYr06}.  The
author's position is that, to the contrary, there are perfectly consistent
ways of applying the basic ideas of classical information theory to small
numbers (even one) of microscopic quantum systems provided attention is paid
to the Hilbert space structure of quantum theory, and probabilities are
introduced in a consistent fashion.  And, further, that this approach has
advantages in that simple systems are simpler to think about than complicated
systems, so it is useful to develop some intuition as to how they behave.  One
goal is to understand both classical and quantum information theory in fully
quantum terms, since the world is (most physicists believe) fundamentally
quantum mechanical.

The basic strategy of this paper is based on the idea that quantum information
comes in a variety of incompatible \emph{types} or \emph{species}. Each type
or species refers to a certain class of (typically microscopic)
mutually-compatible properties of a quantum system.  As long as the discussion
of information is limited to a single type, all the usual formalism and
intuition provided by classical information theory apply directly to the
quantum domain.  On the other hand, incompatible types of quantum information
cannot be combined, as this makes no sense in the context of standard Hilbert
space quantum mechanics.  Since in classical information theory there is only
a single type or species of information, or, equivalently, all different types
are compatible with each other, the main respect in which quantum information
theory needs to go beyond its classical counterpart is in relating
incompatible types of information in a useful way.  Various examples are given
below.

In the real (i.e., quantum) world it must, of course, be the case that
so-called ``classical'' information, as in the acronym LOCC, ``local
operations and \emph{classical} communication,'' is describable in quantum
terms.  A relatively precise definition can be given as indicated in
Sec.~\ref{sct4}: classical information is a particular type of quantum
information, the only one that survives under circumstances (implicitly assumed
in much writing on the subject) where there is strong decoherence.

The remainder of this paper is organized as follows.  The concept of quantum
information types is introduced in the context of a discussion of quantum
incompatibility in Sec.~\ref{sct2}.  The idea is illustrated by the simple
examples of one- and (standard) two-bit teleportation in Sec.~\ref{sct3},
where using quantum information types and what we call the Presence theorem
helps understand why one or two bits, respectively, are needed in these
protocols, or ``dits'' in their $d$-dimensional (``qudit'')
generalizations.
Decoherence and ``classical'' information are the subjects of Sec.~\ref{sct4},
which begins with a simple beam splitter example that illustrates the
importance of the Exclusion theorem, derived from a more general Truncation
theorem, and this sets the stage for a proper understanding of classical
information in quantum terms.

Quantum information theory requires theorems that relate different types of
information, and hence go beyond anything in classical information theory.
Those used in Secs.~\ref{sct3} and \ref{sct4} and some others closely related
to them are given precise formulations in Sec.~\ref{sct5}, extending earlier
work in \cite{Grff05}.  They all have the ``smell'' of no-cloning
\cite{WtZr82}, but the connection is not altogether straightforward, as shown
by an additional Generalized No-Cloning theorem.  Proofs and some additional
technical details are found in the appendices.  A summary and an indication of
various ways the present work needs to be extended comprise the concluding
Sec.~\ref{sct6}.

\section{Types of Information}
\label{sct2}

Central to the following discussion will be the notion of \emph{quantum
  incompatibility} \cite{ntk02}, which can be illustrated using the familiar
two-dimensional Hilbert space of a spin-half particle.  Each one-dimensional
subspace or ray, which is to say all complex multiples of a fixed nonzero ket
$\ket{w}$, is associated with the property that a particular component of
angular momentum is positive, $S_w=+1/2$ (in units of $\hbar$) for some
direction $w$ in space, e.g., $w=z$ or $w=x$ or $w=-z$, etc.  The negation of
the property (or proposition) $S_w=+1/2$ is the property $S_w=-1/2$,
corresponding to the orthogonal complement of the ray associated with
$S_w=+1/2$. In the notation commonly used in quantum information theory,
$S_z=+1/2$ and $-1/2$ correspond to rays passing through (i.e., multiples of)
$\ket{0}$ and $\ket{1}$, respectively, and of course these are orthogonal,
$\inpd{0}{1}=0$.

It always makes sense to talk about the conjunction $P$ AND $Q$ of two
properties $P$ and $Q$ of a classical system, such as $p>0$ and $x <0$ for a
harmonic oscillator. The result may be a property that is always false, as
when $P$ is $p>0$ and $Q$ is $p<0$, and in this case the negation (NOT $P$) OR
(NOT $Q$) of $P$ AND $Q$ is the property $p\leq 0$ OR $p\geq 0$, which is
always true.  But in quantum theory it is possible to write down conjunctions,
such as
\begin{equation}
  S_x=+1/2 \text{ AND } S_z=+1/2,
\label{eqn1}
\end{equation}
which make no sense.  Obviously \eqref{eqn1} cannot correspond to any ray in
the Hilbert space, since each ray is associated with $S_w=+1/2$ for some
direction $w$, and \eqref{eqn1} is not of this form. Could it be a proposition
that is always false?  Then its negation $S_x=-1/2$ OR $S_z=+1/2$ must always
be true, which does not seem very plausible. Indeed, assuming that
\eqref{eqn1} and similar conjunctions are false swiftly leads to a
contradiction if one follows the usual rules of logic---for details, see
Sec.~4.6 of \cite{Grff02c}.  This was understood by Birkhoff and von Neumann
\cite{BrvN36}, who proposed altering the rules of propositional logic to get
around this difficulty.  Their proposal has not been of much use for
interpreting quantum mechanics, which may merely mean that we physicists are
not smart enough.  By contrast, if one restricts the domain of meaningful
discourse so as to exclude \eqref{eqn1} and similar things---in particular,
conjunctions (AND) and disjunctions (OR) of properties corresponding to
projectors that do not commute---it is possible to produce a consistent
interpretation of quantum mechanics
\cite{Grff84,Omns88,GMHr90,GMHr93,Omns94,Grff96,Grff98,Omns99,Grff02c}
 that follows the usual rules of logic (as
applied to meaningful statements), and resolves all the standard paradoxes
\cite{ntk00}.

The compatible propositions $S_z=+1/2$ and $S_z=-1/2$, corresponding to
mutually orthogonal projection operators, form a \emph{quantum sample space} of
mutually exclusive possibilities: their conjunction is always false, and since
each is the negation of the other, one or the other is always true.  This
makes physical sense in that one can in principle carry out a Stern-Gerlach
measurement to determine whether $S_z=+1/2$ or $-1/2$ \cite{ntk03}.  (By
contrast, there is no measurement which can determine the truth or falsity of
\eqref{eqn1}, as one would expect for something that is meaningless.)
Information that answers the question of whether $S_z=+1/2$ or $-1/2$ is what
we shall call the $\ZS$ \emph{type} (or \emph{species}) of information.
Similarly, $\XS$ information answers the question whether $S_x=+1/2$ or
$-1/2$. It is incompatible with $\ZS$ information in that there is no way in
which the two can be meaningfully combined: \eqref{eqn1} makes no sense, and
asking whether $S_z=+1/2$ or $S_x=+1/2$ is equally meaningless.  For a
spin-half particle there is a type of information associated with each pair
$w$ and $-w$ of opposite directions in three-dimensional space, and the
different species associated with distinct pairs are incompatible. 

In larger Hilbert spaces a quantum sample space or type of information always
corresponds to a \emph{decomposition of the identity}, a collection of
mutually orthogonal projectors $\VS=\{V_j\}$, $V_j\ad = V_j = V^2_j$, that sum
to the identity $I$.  In the case of an orthonormal basis, $V_j=\dya{v_j}$ and
$\inpd{v_j}{v_k} = \dl_{jk}$, we also write $\VS = \{\ket{v_j}\}$, since the
meaning is obvious.  Two such collections or types of information $\VS$ and
$\WS$ are \emph{compatible} if and only if all projectors in one commute with
all projectors in the other; otherwise they are \emph{incompatible}. The
``single framework'' rule of quantum reasoning \cite{ntk01} generalizes the
example discussed above, and states that incompatible quantum descriptions
(decompositions, information types) cannot be meaningfully combined.

\section{Teleportation}
\label{sct3}

\begin{figure}[h]
$$
\begin{pspicture}(-1.8,-3.4)(5.0,3.4) 
\newpsobject{showgrid}{psgrid}{subgriddiv=1,griddots=10,gridlabels=2pt}
\def\drad{0.25} 
\def\hdg{0.25}  
\def\lwd{0.025} 
\def\rcp{0.25}  
\def\rdot{0.05} 
\psset{
labelsep=2.0,
arrowsize=0.150 1,linewidth=\lwd}
\def\cnot{\pscircle(0,0){\rcp}\psline(-\rcp,0)(\rcp,0)\psline(0,-\rcp)(0,\rcp)}
\def\detect{\psline(0,-\drad)(0,\drad)\psarc(0,0){\drad}{-90}{90}}
\def\dput(#1)#2#3{\rput(#1){#2}\rput(#1){#3}} 
\def\dot{\pscircle*(0,0){\rdot}} 
\def\odot{\pscircle[fillcolor=white,fillstyle=solid](0,0){\rodot}} 
\def\rectc(#1,#2){%
\psframe[fillcolor=white,fillstyle=solid](-#1,-#2)(#1,#2)}
\def\squ{\psframe[fillcolor=white,fillstyle=solid](-\hdg,-\hdg)(\hdg,\hdg)}
		\def\onebita{%
\psline{>->}(0,0)(4.0,0)
\psline{>-}(0,1)(2,1)
\psline(1,0)(1,1)
\rput(1,1){\dot}
\psline[linestyle=dashed]{->}(2.25,1)(3.0,0.25)
\dput(1,0){\squ}{X}
\dput(1.5,1){\squ}{$H$}
\rput(2,1){\detect}
\dput(3.0,0){\squ}{$Z$}
\rput[b](0.5,1.1){$a$}
\rput[b](0.5,0.1){$b$}
\rput[b](3.6,0.1){$b'$}
\rput[r](-0.1,1){$\ket{\psi}$}
\rput[r](-0.1,0){$\ket{0}$}
\rput[l](4.1,0){$\ket{\psi}$}
\rput[bl](2.7,0.7){$x$}
\rput(-1.5,0.5){\textbf{(a)}}
		}
		\def\onebitb{%
\psline{>->}(0,0)(4.0,0)
\psline{>->}(0,1)(4.0,1)
\psline(1,0)(1,1)
\rput(1,1){\dot}
\psline(3,0)(3,1)
\rput(3,1){\dot}
\dput(1,0){\squ}{X}
\dput(1.5,1){\squ}{$H$}
\dput(3.0,0){\squ}{$Z$}
\rput[b](0.5,1.1){$a$}
\rput[b](3.6,1.1){$a'$}
\rput[b](0.5,0.1){$b$}
\rput[b](3.6,0.1){$b'$}
\rput[r](-0.1,1){$\ket{\psi}$}
\rput[r](-0.1,0){$\ket{0}$}
\rput[l](4.1,0){$\ket{\psi}$}
\rput(-1.5,0.5){\textbf{(b)}}
		}
		\def\onebitc{%
\psline{>->}(0,0)(4.0,0)
\psline{>->}(0,1)(4.0,1)
\psline{>->}(0,2)(4.0,2)
\psline(1,0)(1,1)
\rput(1,1){\dot}
\psline(3,0)(3,1)
\rput(3,1){\dot}
\dput(1,0){\squ}{X}
\dput(1.5,1){\squ}{$H$}
\dput(3.0,0){\squ}{$Z$}
\rput[r](-0.0,1.5){$\lbrk{.5}$}
\rput[l](4.0,1.0){$\rbrk{1.0}$}
\rput[b](0.5,2.1){$\bar a$}
\rput[b](3.6,2.1){$\bar a'$}
\rput[b](0.5,1.1){$a$}
\rput[b](3.6,1.1){$a'$}
\rput[b](0.5,0.1){$b$}
\rput[b](3.6,0.1){$b'$}
\rput[r](-0.5,1.5){$\ket{\Phi}$}
\rput[r](-0.1,0){$\ket{0}$}
\rput[l](4.5,1.0){$\ket{\Psi}$}
\rput(-1.5,0.5){\textbf{(c)}}
		}

\rput(0,0){\onebitb}
\rput(0.0,-3.0){\onebitc}
\rput(0,2.0){\onebita}
\end{pspicture}
$$
\caption{(a) One bit teleportation. (b) Quantized version. (c) Channel ket
  construction.}
\label{fgr1}
\end{figure}
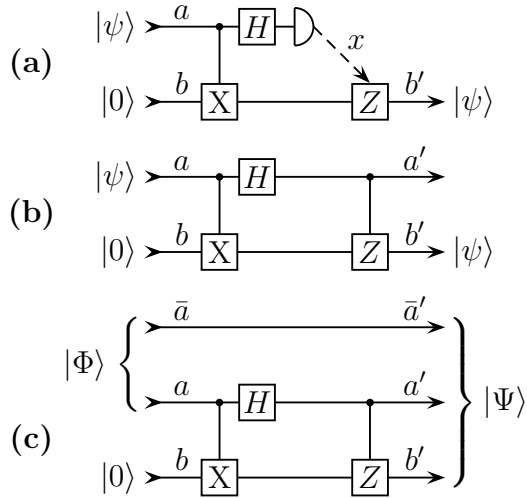

Let us see how using incompatible types of information assists in
understanding why quantum teleportation uses two classical bits of information
in the standard protocol \cite{Bnao93}.  It is simplest to start with a
variant known as ``one bit'' teleportation \cite{ZhLC00}, corresponding to the
quantum circuit in Fig.~\ref{fgr1}(a), where the teleportation process
transports the state $\ket{\psi}$ from the upper left $a$ to the lower right
$b'$.  First, a CNOT, shown as a controlled-X (CX) gate, acts between qubits
$a$ and $b$, and then an $S_x$ measurement is carried out on qubit $a$.  In
the figure this measurement is indicated by the Hadamard gate $H$ that
interchanges $S_x$ and $S_z$, followed by a measurement in the standard or
$S_z$ or ``computational'' basis indicated by the D-like symbol.  If the
measurement reveals $S^a_x=-1/2$ a classical bit (dashed line labeled $x$) is
transmitted to where it actuates a $Z$ gate on qubit $b$, whereas if
$S^a_x=+1/2$ nothing is done. It is an easy exercise to show that whatever
initial state $\ket{\psi}=\al\ket{0}+\bt\ket{1}$ enters at $a$ will later
reappear at $b'$. 

In the case of $\ZS$ information, meaning the input is $\ket{\psi}=\ket{0}$ or
$\ket{1}$, corresponding to $S^a_z=+1/2$ or $-1/2$, the CX (CNOT) gate copies
it from the $a$ to the $b$ qubit so that $S^b_z=S^a_z$, and the later $Z$ gate
has no effect, since even if it acts it only changes the phase of $\ket{1}$,
leaving the ray (or projector) corresponding to $S^b_z=-1/2$ the same.
Thus failing to do the measurement, or throwing away the classical bit, has
no influence so far as transporting the $\ZS$ information is concerned.

In the case of $\XS$ information the input $\ket{\psi}$ is either
$\ket{+}=(\ket{0}+\ket{1})/\st$ or $\ket{-}=(\ket{0}-\ket{1})/\st$,
corresponding to $S_x=+1/2$ or $-1/2$, and the analysis is somewhat more
complicated. The CX gate maps $\ket{\psi}=\ket{+}$ into the two qubit state
$\ket{++} + \ket{--}$ corresponding to $S^b_x=S^a_x$, and $\ket{\psi}=\ket{-}$
into $\ket{+-} + \ket{-+}$, $S^b_x=-S^a_x$. This means the original $\XS$
information is not present in either qubit by itself, since the corresponding
reduced density operator is $\hf I$, but resides in a correlation between the
two.  Information residing in a correlation is not in itself a quantum effect.
One can, for instance, encode a classical bit $\{0_L,1_L\}$ in two coding bits
by letting $00$ or $11$, chosen at random, represent $0_L$, and $01$ or $10$,
again chosen at random, represent $1_L$.  From either coding bit alone one can
extract no information about $0_L$ versus $1_L$, but it is obviously present
in the two together through their correlation.  In the case under discussion
the measurement in Fig.~\ref{fgr1}(a) extracts the value of $S^a_x$
\emph{after} the CX has acted (note that this is \emph{not} the original $\XS$
information), and if this is negative the $Z$ gate applied to qubit $b$
changes the sign of $S^b_x$.  The net effect is that at the end of the process
the value of $S^b_x$ is exactly the same as that of $S^a_x$ at the beginning,
so the $\XS$ information has also been successfully transmitted from $a$ to
$b'$.

One could continue to check what happens to other types of information, but
that is not necessary. The Presence theorem of Sec.~\ref{sct5a} says that once
it is known that \emph{two} suitably incompatible types of information, $\ZS$
and $\XS$ in the case at hand, are present in the output $b'$, all other types
of information about the input $a$ are also present, so there is a perfect
quantum channel from $a$ to $b'$, the desired result for teleportation.  In
summary, the transmission of two (suitably incompatible) types of information
is needed to ensure that there is a good quantum channel from $a$ to $b'$. The
CX gate by itself transmits the $\ZS$ type, while the later measurement and
the single classical bit carrying its outcome are needed to transmit the
incompatible $\XS$ type of information.

The Presence theorem is a statement about \emph{quantum} information discussed
in fully \emph{quantum} terms, so to apply it to the system in
Fig.~\ref{fgr1}(a) one needs to understand the measurement and the
``classical'' bit in quantum terms. This can
be done in the manner indicated in Sec.~\ref{sct4}. But for present purposes
it is convenient to avoid having to introduce the Hilbert space of a
complicated macroscopic system, by ``quantizing'' the circuit in
Fig.~\ref{fgr1}(a) so that it takes the form shown in (b), with the
measurement of the $a$ qubit following the part of the circuit where it
controls (in the usual quantum sense) a $Z$ gate. (See \cite{BrBC98} and
pp.~186f in \cite{NlCh00} for this ``trick,'' based on ideas in
\cite{GrNi96}.) The two circuits in (a) and (b) are equivalent so far as
teleportation is concerned, but the second is simpler to analyze in fully
quantum terms. Indeed, the later measurement of qubit $a$ in (b) need not be
made at all, which is why it is not shown, as its outcome is not used in the
protocol.  (This discussion continues in the latter part of
Sec.~\ref{sct4}.)

In addition, the Presence theorem is stated in Sec.~\ref{sct5a} in the
language of entangled kets, rather than in terms of the input and output of
a quantum channel.  One way of connecting the two is indicated in 
Fig.~\ref{fgr1}(c), where an auxiliary qubit $\bar a$ has been introduced, and
\begin{equation}
  \ket{\Phi} = (\sum_j \ket{a_j}\ot\ket{\bar a_j})/\st,
\label{eqn2}
\end{equation}
where  $\{\ket{a_j}\}$ and  $\{\ket{\bar a_j}\}$ are orthonormal bases
of $\HS_a$ and $\HS_{\bar a}$, is a fully-entangled state.
The result is a final state $\ket{\Psi}\in\HS_{\bar a}\ot\HS_a\ot\HS_b$,
referred to as a ``channel ket'' in \cite{Grff05} (which see for additional
details).  That there is a perfect quantum channel from $a$ to $b'$ in
Fig.~\ref{fgr1}(a) or (b) is the same as saying that all the
information about qubit $\bar a$ is in qubit $b'$ if one uses $\ket{\Psi}$, or
equivalently the reduced density operator obtained by tracing $\dya{\Psi}$
over $\HS_b$, to generate probabilities for correlations between the two in
the usual way.

An alternative way to associate channels with kets is to use \emph{map-state
  duality} \cite{ZcBn04,Grff05} in which an entangled ket
\begin{equation}
  \ket{\psi} = \sum_j\ket{a_j}\ket{\phi_j}
\label{eqn3}
\end{equation}
on the tensor product of $\HS_a$ and another space $\HS_f$ is expanded in an
orthonormal basis of $\HS_a$, with $\{\ket{\phi_j}\}$ the (unnormalized)
expansion coefficients.  One can always ``transpose'' $\ket{\psi}$ to an
operator
\begin{equation}
  M=\sum_j\dyad{\phi_j}{a_j}
\label{eqn4}
\end{equation}
mapping $\HS_a$ to $\HS_f$.  In the particular case in which $\ket{\psi}$ is
maximally entangled, which is to say $\Tr_f(\dya{\psi})$ is proportional to
the identity $I_a$, the map $M$ is, up to normalization, an isometry (a
unitary operator from $\HS_a$ to the subspace $M\HS_a$ of $\HS_f$), that is,
$M\ad M=I_a$.  Conversely, given a map $M$ from $\HS_a$ to $\HS_f$ and an
orthonormal basis of $\HS_a$, it can be expanded in the form \eqref{eqn4}, and
\eqref{eqn3} then defines a corresponding entangled state (typically not
normalized), which when $M$ is an isometry is maximally entangled.  Of course,
if $\ket{\psi}$ is given, $M$ depends on the choice of orthonormal basis
$\{\ket{a_j}\}$, and vice versa, but the Presence theorem is unaffected by the
basis choice. While this and the other theorems in Sec.~\ref{sct5} can be
expressed either in ``map'' or ``entanglement'' language, the latter has the
advantage of being more symmetrical (see remarks in Sec.~VI of \cite{Grff05}).
The idea of regarding the input and output of a quantum channel as
corresponding to the tensor product of two Hilbert spaces, as suggested by the
preceding discussion, is a very natural notion when using atemporal diagrams
\cite{GWYC06}, and within the consistent histories approach to probabilistic
time development \cite{Grff02c}, where the idea of such a tensor product goes
back to Isham \cite{Ishm94}.

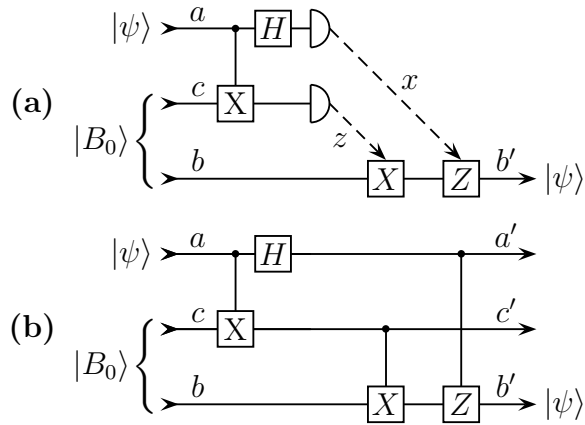
\begin{figure}[h]
$$
\begin{pspicture}(-2.0,-0.4)(5.6,5.4) 
\newpsobject{showgrid}{psgrid}{subgriddiv=1,griddots=10,gridlabels=2pt}
\def\drad{0.25} 
\def\hdg{0.25}  
\def\lwd{0.025} 
\def\rcp{0.25}  
\def\rdot{0.05} 
\psset{
labelsep=2.0,
arrowsize=0.150 1,linewidth=\lwd}
\def\cnot{\pscircle(0,0){\rcp}\psline(-\rcp,0)(\rcp,0)\psline(0,-\rcp)(0,\rcp)}
\def\detect{\psline(0,-\drad)(0,\drad)\psarc(0,0){\drad}{-90}{90}}
\def\dput(#1)#2#3{\rput(#1){#2}\rput(#1){#3}} 
\def\dot{\pscircle*(0,0){\rdot}} 
\def\odot{\pscircle[fillcolor=white,fillstyle=solid](0,0){\rodot}} 
\def\rectc(#1,#2){%
\psframe[fillcolor=white,fillstyle=solid](-#1,-#2)(#1,#2)}
\def\squ{\psframe[fillcolor=white,fillstyle=solid](-\hdg,-\hdg)(\hdg,\hdg)}
		\def\twobita{%
\psline{>->}(0,0)(5.0,0)
\psline{>-}(0,1)(2,1)
\psline{>-}(0,2)(2,2)
\psline(1.0,2)(1.0,1)
\rput(1.0,2){\dot}
\psline[linestyle=dashed]{->}(2.25,1)(3.0,0.25)
\psline[linestyle=dashed]{->}(2.25,2)(4.0,0.25)
\dput(1.0,1){\squ}{X}
\dput(1.5,2){\squ}{$H$}
\rput(2.0,1){\detect}
\rput(2.0,2){\detect}
\dput(3.0,0){\squ}{$X$}
\dput(4.0,0){\squ}{$Z$}
\rput[r](0,0.5){$\lbrk{0.5}$}
\rput[r](-0.4,0.5){$\ket{B_0}$}
\rput[r](-0.1,2){$\ket{\psi}$}
\rput[l](5.1,0){$\ket{\psi}$}
\rput[b](0.5,2.1){$a$}
\rput[b](0.5,0.1){$b$}
\rput[b](4.6,0.1){$b'$}
\rput[b](0.5,1.1){$c$}
\rput[bl](3.2,1.2){$x$}
\rput[tr](2.5,0.6){$z$}
\rput(-1.7,1){\textbf{(a)}}
		}
		\def\twobitb{%
\psline{>->}(0,0)(5.0,0)
\psline{>->}(0,1)(5.0,1)
\psline{>->}(0,2)(5.0,2)
\psline{>-}(0,1)(2,1)
\psline{>-}(0,2)(2,2)
\psline(1.0,2)(1.0,1)
\rput(1.0,2){\dot}
\psline(3.0,1)(3.0,0)
\rput(3.0,1){\dot}
\psline(4.0,2)(4.0,0)
\rput(4.0,2){\dot}
\dput(1.0,1){\squ}{X}
\dput(1.5,2){\squ}{$H$}
\dput(3.0,0){\squ}{$X$}
\dput(4.0,0){\squ}{$Z$}
\rput[r](0,0.5){$\lbrk{0.5}$}
\rput[r](-0.4,0.5){$\ket{B_0}$}
\rput[r](-0.1,2){$\ket{\psi}$}
\rput[l](5.1,0){$\ket{\psi}$}
\rput[b](0.5,2.1){$a$}
\rput[b](4.6,2.1){$a'$}
\rput[b](0.5,0.1){$b$}
\rput[b](4.6,0.1){$b'$}
\rput[b](0.5,1.1){$c$}
\rput[b](4.6,1.1){$c'$}
\rput(-1.7,1){\textbf{(b)}}
		}
\rput(0,0){\twobitb}
\rput(0,3.0){\twobita}
\end{pspicture}
$$
\caption{(a) Standard (two bit) teleportation. (b) Quantized version.}
\label{fgr2}
\end{figure}

Conventional ``two bit'' teleportation, Fig.~\ref{fgr2}, with
$\sqrt{2}\ket{B_0}=\ket{00}+\ket{11}$, can be analyzed in the same way; the
details are left as an exercise. The two classical bits in (a) are labeled $x$
and $z$ to indicate that they are essential for correct transmission of the
$\ZS$ and $\XS$ information, respectively; throwing $z$ away will not affect
$\XS$ information, and $x$ is dispensable if only $\ZS$ information is of
interest.  Neither classical bit, nor the two together, actually contain any
information in themselves, a consequence of No Splitting, Sec.~\ref{sct5b}.
All the information is in correlations of the classical bits with the $b$
qubit (more details in\cite{Grff02}).  One classical bit is needed for each of
these two incompatible types of quantum information, but that is enough, for
the Presence theorem then guarantees that all other species are correctly
transmitted, so one has a perfect quantum channel from $a$ to $b'$.  Figure
2(b) is the quantized version of (a), which is convenient for analyzing
the situation in quantum terms, and one could once again introduce a channel
ket $\ket{\Psi}$ (not shown), this time on four qubits, in analogy with
Fig.~\ref{fgr1}(c). The Presence theorem can be applied to the channel ket, or
one can regard the channel entrance as constituting its own Hilbert space.
Generalization to the teleportation of a qudit (with Hilbert space of
dimension $d>2$) is straightforward: if two suitably incompatible types of
information are correctly transmitted---each type requires a classical
``dit''---one has a perfect quantum channel.

\section{Decoherence and Classical Information}
\label{sct4}
 
\begin{figure}[h]
$$
\begin{pspicture}(-0.75,-1.2)(3.75,1.2) 
\newpsobject{showgrid}{psgrid}{subgriddiv=1,griddots=10,gridlabels=6pt}
\def\lwd{0.025} 
\def\lwb{0.10}  
\def\mirw{0.4}\def\mirt{0.1}  
\psset{
labelsep=2.0,
arrowsize=0.150 1,linewidth=\lwd}
\def\dput(#1)#2#3{\rput(#1){#2}\rput(#1){#3}} 
\def\rectg(#1,#2,#3,#4){
\psframe[fillcolor=white,fillstyle=solid](#1,#2)(#3,#4)}
\def\mirra{\rectg(-\mirw,0,\mirw,\mirt)}
\def\mirrb{\rectg(-\mirw,-\mirt,\mirw,0)}
\def\bspl{\psline[linewidth=0.1,linestyle=dashed](-0.3,0)(0.3,0)}
\psline(-0.75,0.5)(1.5,-1)(3,0)
\psline{->}(-0.75,0.5)(-0.45,0.3)
\psline{->}(0,0)(1.5,1)(3.75,-0.5)
\psline[linestyle=dashed]{->}(3,0)(3.75,0.5)
\rput(0,0){\bspl}\rput(3,0){\bspl}
\rput(1.5,1){\mirra}\rput(1.5,-1){\mirrb}
\rput[r](-0.4,0){$B$}
\rput[r](2.6,0){$B'$}
\rput[b](-0.25,0.35){$d$}
\rput[b](0.75,0.65){$e$}
\rput[b](2.25,0.65){$e$}
\rput[t](0.75,-0.65){$f$}
\rput[t](2.25,-0.65){$f$}
\rput[b](3.25,0.35){$g$}
\rput[t](3.25,-0.35){$h$}
\end{pspicture}
$$
\caption{Interferometer with beamsplitters $B$ and $B'$.}
\label{fgr3}
\end{figure}
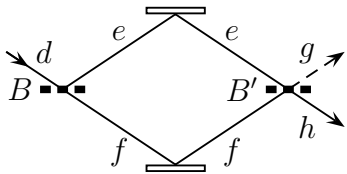

An application of types of information to a simple case of decoherence is
shown in Fig.~\ref{fgr3}, where a particle (neutron or photon) enters an
interferometer on path $d$ at beamsplitter $B$ and, because at an intermediate
time it is in a coherent superposition $(\ket{e}+\ket{f})/\st$, leaves the
second beamsplitter $B'$ in channel $h$ rather than $g$.  But if while inside
the interferometer some interaction with the environment leaves a trace
indicating that the particle took path $e$ rather than $f$, or vice versa, the
interference effect is lost, and the particle emerges with equal probability
in $g$ or $h$.  Let $\ZS$ be the $e$ vs.\ $f$ ``which way'' information, and
$\XS$ be the $(\ket{e}+\ket{f})/\st$ vs.\ $(\ket{e}-\ket{f})/\st$ ``coherent''
information. Decoherence, the disappearance of coherence, in this case $\XS$
information, when $\ZS$ information about the path resides in the environment,
illustrates the Exclusion theorem of Sec.~\ref{sct5}: one type of information
about $\ST_a$ perfectly present in $\ST_b$ means that a mutually-unbiased type
is completely absent from $\ST_c$.  Two types of information $\XS$ and $\ZS$
are said to be mutually unbiased if they correspond to mutually unbiased
orthonormal bases $\{\ket{x_j}\}$ and $\{\ket{z_k}\}$, with
$|\inpd{x_j}{z_k}|^2$ equal to 1 divided by the dimension of the Hilbert
space, independent of $j$ and $k$.

To apply this theorem to the situation in Fig.~\ref{fgr3}, think of the
particle that has just passed through the first beam splitter as system
$\ST_a$, and just before it reaches the second beam splitter $B'$ as $\ST_c$,
while $\ST_b$ is the environment at this second time.  (See the discussion in
Sec.~\ref{sct3} on why one can regard the particle at two different times as
two separate systems, and how to apply the Exclusion theorem, worded in terms
of entangled states, to situations with unitary time evolution.)  For our
purposes it suffices to model $\ST_a$ and $\ST_c$ using a $d=2$ dimensional
Hilbert space spanned by $\ket{e}$ and $\ket{f}$---this is analogous to
focusing on the spin of a particle when its other degrees of freedom are not
relevant to the analysis.
The Exclusion theorem says that when the $\ZS$ or which-way information about
$\ST_a$ is perfectly present in the environment, i.e., at the time the particle
reaches the second beam splitter, the (mutually unbiased) $\XS$ or coherence
information must be perfectly absent from $\ST_c$, i.e., from the particle
itself at this later time.  And in the absence of coherence all interference
effects disappear: the situation after the second beam splitter is,
statistically, just the same as if the particle arrived at random on path $e$
or path $f$.  All this is well known, and the connection between decoherence
and information in the environment has been previously pointed out by Zurek
and his collaborators \cite{Zrk03,OlPZ04,BKZr05,OlPZ05}.  The use of
\emph{types} of information, not tied to some notion of measurement
\cite{ntk05}, is our attempt to add further clarity and precision to these
seminal ideas.

The situation to which the Exclusion theorem applies is that of \emph{strong},
meaning essentially complete, decoherence.  Clearly extensions are needed
(Sec.~\ref{sct6}) to cases of only partial decoherence.  Nevertheless, strong
decoherence is a useful idealization both because if is often a good
approximation to what is realized in the laboratory (to the dismay of
those who want to build quantum computers), and because it yields a precise
definition of another idealization, \emph{classical information}.
Indeed, it is rather odd to find the term ``classical information'' floating
around in technical books and articles on quantum information theory when
most, even if not all, physicists believe that all physical processes in the
real world are quantum mechanical, with classical physics a good approximation
in appropriate circumstances, but hardly part of our fundamental understanding
of nature.  A good way to see how ``classical'' information can arise in
quantum mechanics is to note that one consequence of the Truncation theorem as
discussed in Sec.~\ref{sct5b} is the fact that if a particular type of
information about $\ST_a$ associated with an orthonormal basis $\{\ket{v_j}\}$
is perfectly present in $\ST_b$ it is the \emph{only} type of information
about $\ST_a$ which can be present in a third system $\ST_c$ in the sense that
any other species of information is parasitic upon, or controlled by, or
compatible with the $\{\ket{v_j}\}$ type.  Whenever \emph{only one} type of
information needs to be considered all the rules of classical information
theory apply to it; conversely, ``classical information'' in the quantum
context refers to the single dominant type of \emph{quantum} information
available in a situation of strong decoherence.  Typically it is the presence
of this type of information in the environment that means that other types can
be ignored in systems which are not isolated from the environment.  In
particular, the measurements indicated in part (a) of Figs.~\ref{fgr1} and
\ref{fgr2} when instantiated in physical apparatus amplify a particular type
of information, and the environment rapidly copies the ``pointer positions,''
resulting in strong decoherence.  To avoid the rather unwieldy task of trying
to describe this amplification process and interaction with the environment in
correct quantum mechanical terms, which is certainly possible in principle, it
is often preferable (as noted earlier) to employ a simple quantum circuit in
which the decoherence is ``built in'': the $a$ qubit in Fig.~\ref{fgr1}(b),
and the $a$ and $c$ in qubits in Fig.~\ref{fgr2}(b), are at later times good
copies of the $\ZS$ information preceding the final control gates, and since
no further use is made of them, they may be regarded as carrying this type of
information off into the environment.

\section{Theorems}
\label{sct5}

In this section we state and prove results used in the preceding sections,
plus some additional ones that are closely related.  The treatment builds upon
ideas and terminology from \cite{Grff05}, repeated here to the extent needed
to make the exposition self-contained.  Note in particular that $\HS_a$ is the
Hilbert space of system $\ST_a$, $\HS_{ab}=\HS_a\ot\HS_b$ that of $\ST_{ab}$,
the systems $\ST_a$ and $\ST_b$ regarded as a single system, $\rho_{ab}$ is a
density operator on $\HS_{ab}$, often traced down from that of a larger
system, $d_a$ the dimension of $\HS_a$, and so forth. All Hilbert spaces are
assumed to be of finite dimension in order to avoid technical complications. 

\subsection{Presence}
\label{sct5a} 

\Thm (Presence). \emph{
  Let $\ST_a$ and $\ST_b$ be two quantum systems with Hilbert spaces $\HS_a$
  and $\HS_b$, and $\VS=\{V_j\}$ and $\WS=\{W_k\}$ two strongly incompatible
  projective decompositions of the identity $I_a$.  If both the $\VS$ and the
  $\WS$ information is perfectly present in $\ST_b$ for a density operator
  $\rho$ on $\HS_{ab}$ (possibly a pure state), then \emph{all} types of
  information about $\ST_a$ are perfectly present in $\ST_b$.
}

The terms are to be understood as follows.  The density operator $\rho$ on
$\HS_{ab}$ or pure state $\ket{\Psi}\in\HS_{ab}$ will be called a
\emph{pre-probability} using the terminology of Ch.~9 of \cite{Grff02c},
because it can be used to generate probabilities once a quantum sample
space---an orthonormal basis of $\HS_{ab}$ or a decomposition of the identity
$I_{ab}$---has been specified, following the usual rule that the probability
associated with a projector $P$ is
\begin{equation}
  \Pr(P) =  \lgl P\rgl = \Tr(P\rho) = \mte{\Psi}{P},
\label{eqn5}
\end{equation}
with $\rho=\dya{\Psi}$ for a pure state.  For example, $\rho$ as a density
\emph{operator} is represented by different \emph{matrices} if different
orthonormal bases are chosen.  The diagonal elements of one of these matrices
form a probability distribution associated with the corresponding basis (or
type of information), whereas the single density operator giving rise to the
different distributions is the pre-probability.

The $\VS$ type of information is \emph{perfectly present} in $\ST_b$ for a
given pre-probability (Sec.~III C of \cite{Grff05}) when the unnormalized
conditional density operators
\begin{equation}
  \rho_{bj} = \Tr_a\lp V_j \rho\rp
\label{eqn6}
\end{equation}
on $\HS_b$ are mutually orthogonal, i.e.,
\begin{equation}
 \rho_{bj}\,\rho_{bk}=0 \text{ for }j\neq k.
\label{eqn7}
\end{equation}
In the language of measurements, if one thinks of carrying out a projective
measurement on $\HS_a$ corresponding to $\{V_j\}$, then there is a
corresponding decomposition $\{T_k\}$ of $I_b$ such that the measurement
outcomes are in one-to-one correspondence.  An analogous definition applies to
$\WS$ information.  The conclusion of the theorem, that all species of
information about $\ST_a$ are perfectly present in $\ST_b$, conveniently
abbreviated to ``all information about $\ST_a$ is in $\ST_b$,'' means that for
any decomposition of the identity, in particular for any orthonormal basis of
$\HS_a$, that kind of information is perfectly present in $\ST_b$ in the sense
just discussed.  When the pre-probability is a pure state $\ket{\Psi}$, this
implies it is maximally entangled, i.e., $\rho_a=\Tr_b(\dya{\Psi})$ is
proportional to the identity $I_a$.  For a general $\rho$ a similar but more
complicated result obtains---see theorem 3(ii) in \cite{Grff05}---and once
again $\rho_a=\Tr_b(\rho)$ is proportional to $I_a$.

The decompositions $\VS$ and $\WS$ are said to be \emph{strongly incompatible}
(Sec.~IV of \cite{Grff05}) when the only projector $P$ that commutes with
\emph{every} $V_j$ and \emph{every} $W_k$ is either $P=0$ or $P=I_a$.  While
concise, this definition is not very intuitive.  In the case of orthonormal
bases $\VS=\{\ket{v_j}\}$ and $\WS=\{\ket{w_k}\}$ one can use a somewhat
simpler definition. Construct a graph containing $2d_a$
nodes, one for each $\ket{v_j}$ and one for each $\ket{w_k}$. Whenever the
inner product $\inpd{v_j}{w_k}$ is nonzero, draw an edge between the
corresponding nodes.  Then $\VS$ and $\WS$ (i.e., the corresponding
collections of projectors) are strongly incompatible if and only if this graph
is \emph{connected}.  The proof is at the end of App.~\ref{sctpa}.
In the case of two mutually unbiased bases, as in Sec.~\ref{sct4}, every
$\{\ket{v_j}\}$ is connected to every $\{\ket{w_k}\}$ node, so connectivity of
the graph is obvious.  It is equally obvious for two bases in which
$\inpd{v_j}{w_k}$ is never zero.  However, strong incompatibility can still
hold if some of the $\inpd{v_j}{w_k}$ are zero, provided the graph remains
connected.

The proof of the Presence theorem, extending a weaker theorem in \cite{Grff05},
is in App.~\ref{sctpa}

\subsection{Truncation, Exclusion, No Splitting, Somewhere}
\label{sct5b}

A series of useful ``all-or-nothing'' results about information in three
systems $\ST_a$, $\ST_b$, and $\ST_c$ begins with:

\Thm (Truncation). \emph{
  Let $\ST_a$, $\ST_b$ and $\ST_c$ be three quantum systems, and suppose that
  for some decomposition $\VS=\{V_j\}$ of $I_a$ and for some density operator
  $\rho$ on $\HS_{abc}$ all the $\VS$ information about $\ST_a$ is present in
  $\ST_b$.  Then any other type of information $\WS=\{W_k\}$ about $\ST_a$
  will be ``truncated'' (or ``censored'') in the sense that
\begin{equation}
  \rho_{ac} = \sum_j V_j \rho_{ac} V_j,
\label{eqn8}
\end{equation}
that is, $\rho_{ac}$, the partial trace of $\rho$ over $\HS_b$, commutes with
all the $V_j$. (Note that $V_j$ is here understood as $V_j\ot I_c$ on
$\HS_{ac}$.) Equivalently, all correlations between $\ST_a$ and the third
system $\ST_c$ satisfy
\begin{equation}
  \lgl AC\rgl = \lgl \bar A C\rgl
\label{eqn9}
\end{equation}
for any operators $A$ and $C$ on $\HS_a$ ad $\HS_c$, respectively (one
could write $A\ot C$ in place of $AC$), with 
\begin{equation}
  \bar A = \sum_j V_j A V_j. 
\label{eqn10}
\end{equation}
the truncated version of the operator $A$, and $\lgl\;\rgl$ the average taken
with respect to $\rho$, as in \eqref{eqn5}.
}

This theorem is closely related to, but not the same as, theorem 6(i) in
\cite{Grff05}, and its proof is in App.~\ref{sctpb}.
Since any operator $A$ can be written as $A=\sum_{jk} V_j A V_k$, its
truncated version $\bar A$ is obtained by throwing away the off-diagonal
blocks. To understand the implications of the theorem it helps to consider the
case in which $\VS=\{\ket{v_j}\}$ is an orthonormal basis of $\HS_a$, so that
\begin{equation}
  \bar A = \sum_j \mte{v_j}{A} \dya{v_j}
\label{eqn11}
\end{equation}
is diagonal in this basis, meaning that all correlations between $A$ and $C$
can be computed from the correlations $\lgl V_j C\rgl$, that is from $\VS$
information about $\ST_a$ in $\ST_c$. Equivalently, 
\begin{equation}
  \rho_{ac} = \sum_j\dya{v_j}\ot \Gm_j,
\label{eqn12}
\end{equation}
where the $\Gm_j$ are operators on $\HS_c$.  All other information about
$\ST_a$ in $\ST_c$, of whatever kind, is then ``parasitic upon,'' ``truncated
by,'' or ``censored relative to'' the $\VS$ information.  When the $V_j$
projectors have rank greater than 1 the truncation or censorship is less
extreme, but it remains true that the only sort of information about $\ST_a$
allowed in $\ST_c$ is represented by $\bar A$-type operators which are
compatible with $\VS$ in the sense of commuting with every $V_j$ or,
equivalently, $\rho_{ac}$ commutes with every $V_j$. 

The situation is particularly clear if there is another basis
$\WS=\{\ket{w_k}\}$ which is mutually unbiased with respect to
$\VS=\{\ket{v_j}\}$, i.e., $|\inpd{w_k}{v_j}|^2=1/d_a$ for all $j$ and $k$. In
that case the truncated projectors $\bar W_k$ are not only diagonal in the
$\VS$ representation, but are all equal to $I_a/d_a$ independent of $k$,
proving the next theorem (which is the same as theorem 7(ii) in
\cite{Grff05}):

\Thm (Exclusion). \emph{
  Let $\ST_a$, $\ST_b$ and $\ST_c$ be three quantum systems, and
  $\VS=\{\ket{v_j}\}$ and $\WS=\{\ket{w_k}\}$ two mutually unbiased
  orthonormal bases of $\HS_a$.  Then if the $\VS$ information about $\ST_a$
  is perfectly present in $\ST_b$, the $\WS$ information about $\ST_a$ is
  perfectly absent from $\ST_c$.
} 

The \emph{perfect absence} of some type of information can be defined using
reduced density operators, as in \eqref{eqn6}, but now they are required to be
the same up to a multiplicative constant.  That is, the $\WS$ or $\{W_k\}$
information about $\ST_a$ is perfectly absent from $\ST_c$ if and only if for
every $k$
\begin{equation}
  \Tr_a\lp W_k \rho_{ac}\rp = p_k\rho_c,
\label{eqn13}
\end{equation}
where $\rho_c=\Tr_a(\rho_{ac})$, $\rho_{ac}$ is the density operator
(pre-probability) on $\HS_{ac}$, and the $p_k$ are nonnegative numbers.  One
can think of this in terms of measurements as saying that when a projective
measurement corresponding to $\{W_k\}$ is made on $\ST_a$, the probability of
any measurement on $\ST_c$ conditioned on the outcome $k$ will be independent
of $k$. 

Below we will need the notion of the perfect absence of \emph{all} types of
information about $\ST_a$ from $\ST_c$, conveniently abbreviated to ``no
information about $\ST_a$ is in $\ST_c$.'' This is equivalent to
$\rho_{ac}=\rho_a\ot\rho_c$, or to $\ket{\Psi}=\ket{\al}\ot\ket{\gm}$ for a
pure state, theorem 1(iii) of \cite{Grff05}. As the relationship is obviously
symmetrical, one can also say that $\ST_a$ and $\ST_c$ are
\emph{uncorrelated}.

An important corollary of the Exclusion theorem is:

\Thm (No Splitting). \emph{
  Let $\ST_a$, $\ST_b$ and $\ST_c$ be three quantum systems.  If all types of
  information about $\ST_a$ are perfectly present in $\ST_b$, then all types
  will be perfectly absent from $\ST_c$. That is, if all information about
  $\ST_a$ is in $\ST_b$, no information about $\ST_a$ is in $\ST_c$.
}

This is theorem 8(i) in \cite{Grff05}. It follows at once from the Exclusion
theorem, because to show the absence of some species of  information
about $\ST_a$ in $\ST_c$ it suffices to consider orthonormal bases, 
and for each of these we know that there is at least one mutually unbiased
basis for which all the corresponding information is, by hypothesis, 
perfectly present in $\ST_b$.
The No Splitting theorem has lots of applications.  For example, in either one
or two bit teleportation after the final corrections have been made, there is
no information about the input state $\ket{\psi}$ remaining in the environment
treated as a quantum system, and since copies of the classical bits $x$ and
$y$ used to complete the protocol can remain in the environment, it is evident
that they, as has often been observed, can contain no information about the
input: their probabilities cannot depend upon $\ket{\psi}$.  In the case of
quantum codes the presence of the encoded information in some subset of the
coding bits (which is what makes error correction possible) means its absence
from the complementary subset of coding bits, and this can provide additional
intuition about the coding process \cite{Grff05}.

Is there a converse to the Exclusion theorem which says that if the $\WS$
information about $\ST_a$ is perfectly absent from $\ST_c$, then that
associated with any mutually unbiased basis $\VS$ must be present in $\ST_b$?
No, not even if one knows that all information about $\ST_a$ is present in the
combined system of $\ST_{bc}$; see the end of App.~\ref{sctpb} for a
counterexample.  There is, on the other hand, a partial converse of the
No Splitting theorem:

\Thm (Somewhere). \emph{
  If for a \emph{pure state} pre-probability $\ket{\Psi}$ on $\HS_{abc}$ it is
  the case that all the information about $\ST_a$ is in the combined system
  $\ST_{bc}$, and none of it is in $\ST_c$, then it is all in $\ST_b$.
} 

The name ``Somewhere'' comes from the idea that if we know that an object is
in one of two rooms and it is not in the second, it has to be in the first: it
must be \emph{somewhere}.  However, information is very different from a lost
child, as it can be present in \emph{correlations} between two systems, while
not being available in either system by itself.  See, for example, the
discussion of $\XS$ information in one bit teleportation in Sec.~\ref{sct3}.
Consequently, the Somewhere theorem is a decidedly quantum mechanical result.
Also, it fails (in general) if the pure state is replaced by a density
operator.  The theorem itself is proved in \cite{Grff05} as theorem 8(ii),
where one will also find an application to quantum codes.

\subsection{Absence}
\label{sct5c}  

Given the Presence theorem one might anticipate a similar Absence
theorem.  It comes in two versions:

\Thm (Absence). \emph{
Let $\ST_a$ and $\ST_b$ be two quantum systems.
}

\emph{
i) Simple version.  If the pre-probability is a \emph{pure} state $\ket{\Psi}$
on $\HS_a\ot\HS_b$ and the information associated with a single orthonormal
basis $\{\ket{v_j}\}$ of $\HS_a$ is completely absent from $\HS_b$, then
$\ket{\Psi}$ is a product state of the form $\ket{a}\ot\ket{b}$, so
there is no information about $\ST_a$ in $\ST_b$ or vice versa; the two are
uncorrelated. 
}

\emph{
ii) Complicated version.  Let the pre-probability be a general density
operator on $\HS_a\ot\HS_b$, and let $\{V^{(m)}\}$ be a collection of
decompositions of the identity $I_a$ of $\HS_a$,
\begin{equation}
 I_a = \sum_j V^{(m)}_j  
\label{eqn14}
\end{equation}
for each $m$, where the $V^{(m)}_j$ for different $j$ are the projectors
belonging to $V^{(m)}$.  If the collection $\{V^{(m)}_j\}$ for all $m$ and all
$j$ spans the space $\hat\HS_a$ of operators on $\HS_a$, and if each species
$V^{(m)}$ of information about $\ST_a$ is completely absent from $\ST_b$, then
\begin{equation}
  \rho = \rho_a \ot \rho_b,
\label{eqn15}
\end{equation}
so there is no information about $\ST_a$ in $\ST_b$ or vice versa; the two
are uncorrelated.
}

The proof of both versions is given in App.~\ref{sctpc}.  For version (ii) the
conditions are definitely more complicated than for the Presence theorem: if
$\HS_a$ has dimension $d_a$, one needs to check not two but at least $d_a+1$
orthonormal bases (see end of App.~\ref{sctpc}) in order to be sure that all
information is absent.  For instance, if $\ST_a$ is a qubit, $d_a=2$, it
suffices to check that the $\XS$, $\YS$, and $\ZS$ types of information are
absent, but two out of the three is not enough, as shown in the example at the
end of App.~\ref{sctpb}.

\subsection{No Cloning}
\label{sct5d}

One might suspect that the No-Splitting theorem is the same as, or at least
closely related to, the well-known no cloning result \cite{WtZr82}.  However,
the two seem to be different, since neither the conditions nor the
consequences of no-cloning are expressed in terms of types of information as
used here.  The following theorem is the closest we have been able to come in
finding a connection between the two.

\Thm (Generalized No Cloning). \emph{
  Let $M$ be an isometry from $\HS_a$ to $\HS_{bc}$, and $\{\ket{\al_j}\}$,
  with $j$ lying in a finite index set $J$, a collection of normalized kets on
  $\HS_a$ with the property that the pairs $(j,k)$ for which the inner product
  $\inpd{\al_j}{\al_k}$ is nonzero when treated as edges produce a connected
  graph on the set $J$.  Assume that for each $\ket{\al_j}$ its image under
  $M$ is a product state
\begin{equation}
  M \ket{\al_j} = \ket{\bt_j}\ot\ket{\gm_j},
\label{eqn16}
\end{equation}
where both $\ket{\bt_j}$ and $\ket{\gm_j}$ are normalized, and that
\begin{equation}
  |\inpd{\al_j}{\al_k}| = |\inpd{\bt_j}{\bt_k}|
\label{eqn17}
\end{equation}
whenever the left side is nonzero.
}

\emph{
Under these conditions $M$ restricted to the subspace $\GS_a$ spanned by the
$\{\ket{\al_j}\}$ is of the form
\begin{equation}
  M\ket{a} = \lp U\ket{a}\rp\ot \ket{\gm_1},
\label{eqn18}
\end{equation}
where $U$ is a unitary map of $\GS_a$ onto the subspace 
$\GS_b$ of $\HS_b$ spanned by the $\{\ket{\bt_j}\}$, and $\ket{\gm_1}$ is a
fixed ket in $\HS_c$. 
}

The proof is in App.~\ref{sctpd}.  The connection with no-cloning, not obvious
given the somewhat abstract statement of the theorem, is the following.
Suppose $j$ takes on just two values 1 and 2, the states $\ket{\al_1}$ and
$\ket{\al_2}$ are linearly independent, and $\inpd{\al_1}{\al_2}\neq 0$.
Imagine these are two states to be cloned, and the isometry $M$ (which can be
replaced with a unitary acting on the tensor product of $\HS_a$ and an
additional space $\HS_s$ initially in a state $\ket{s_0}$) is supposed to
carry out the cloning process.  If $\ket{\bt_1}$ and $\ket{\bt_2}$ are good
copies up to some unitary transformation of $\HS_b$, their inner product must
equal $\inpd{\al_1}{\al_2}$ apart from an unimportant phase.  As the
conditions of the theorem are fulfilled---the graph consists of two nodes
joined with the edge (1,2)---it follows that $M$ is not only unable to
produce additional copies in $\GS_c$, but in fact there is no information at
all in $\GS_c$ which would allow distinguishing the states $\ket{\al_1}$ and
$\ket{\al_2}$.  Thus at least for the subspace $\GS_a$ (which could be all of
$\HS_a$ if the span of $\{\ket{\al_j}\}$ is large enough) one arrives at the
same conclusion as with the No Splitting theorem, but using somewhat different
hypotheses.

\section{Conclusion}
\label{sct6}

Identifying types or species of quantum information and noting when they are
compatible (i.e., the projectors commute) or incompatible looks like a
promising approach to the foundations of quantum information for the following
reasons. First, it allows a more intuitive, as well as a fully consistent,
approach to quantum probabilities at the microscopic level, in contrast to the
usual textbook approach, with its preparations, measurements, and ``great
smoky dragon'' \cite{Whlr83}, long known to provide an awkward and difficult
(and internally inconsistent \cite{Wgnr63,Mttl98}) way of thinking about the
quantum world, however effective it may be as a calculational tool for the
final outcomes of measurements.  Second, the ideas of classical information
theory \cite{CvTh91} are directly applicable to quantum systems as long as one
restricts oneself to a \emph{single} type of quantum information, or to two or
more \emph{compatible} types (which can then be combined to form a single
type), because there is a properly defined sample space on which probabilities
of \emph{quantum} events and processes, and their correlations, satisfy the
standard rules of probability theory, which are fundamental to the structure
of information theory as developed by Shannon and his successors.  Note that
it is not necessary to restrict oneself to macroscopic systems or
asymptotically large $N$ (number of transmissions, or whatever) limits.

Third, the existence of \emph{different} incompatible species of quantum
information is at the heart of the objections raised in \cite{BrZl01} to
extending Shannon's theory to the quantum domain. Recognizing the role of
quantum incompatibility and using different information types gets around
these problems and allows a fully consistent formulation of the microscopic
statistical correlations needed to properly begin the ``quantization'' of
classical information theory.  Fourth, one of the principal ways quantum
information goes beyond its classical counterpart is in its discussion of how
incompatible types of information relate to, or so-to-speak constrain, each
other for a given setup, or quantum circuit, or entangled state.  The
Presence, Truncation, and Absence theorems and their various corollaries in
Sec.~\ref{sct5} clearly do not belong to the domain of classical information,
since their very formulation requires reference to noncommuting operators, the
hallmark of ``quantum'' effects.

The approach presented here provides, we believe, new intuitive insight into
the processes of teleportation and decoherence, and into how ``classical''
information can be consistently described as a quantum phenomenon.  It is
obviously incomplete in two respects.  First, the theorems of Sec.~\ref{sct5}
are of the ``all or nothing'' variety: they apply to extreme situations in
which information is either completely present or completely absent.
Obviously it would be valuable to have quantitative extensions of these
theorems, presumably in the form of inequalities, that apply to situations
where information of different kinds is partially present or absent.  Finding
suitable information measures and proving appropriate bounds looks like a
challenging problem, but one that needs to be addressed given that one is
often interested in situations where there is noise, so different types of
quantum information will be degraded in different ways.  There are, of course,
many inequalities involving quantum information in the published literature,
and some of them, such as those of Hall \cite{Hll95,Hll97}, look as if they
can be reformulated to apply to different species of information as discussed
here.

Second, the examples and theorems given in this paper (and their extensions
beyond ``all or nothing'' noted above) need to be generalized to cases in
which microscopic quantum properties are considered at a large number of
successive times, as in the case of ``quantum jumps''
\cite{PlKn98,WsTm99,Sdbr02}.  For this purpose it is likely that the full
machinery of quantum histories \cite{Grff02c} will be needed in order to
provide consistent probabilistic descriptions without having to invoke the
awkward concepts of macroscopic ``preparation'' and ``measurement,'' which are
obviously not a fundamental part of microscopic quantum mechanics.

\section*{Acknowledgments}

Helpful comments by Scott M. Cohen and suggestions by an anonymous referee are
gratefully acknowledged.  The research described here received support from
the National Science Foundation through Grant PHY-0456951.

        \appendix
\numberwithin{equation}{section}
\section{Presence Theorem}
\label{sctpa}

The Presence theorem of Sec.~\ref{sct5} is an extension of theorem 4 of
\cite{Grff05}, where it was shown to hold when $\rho$ is a pure state on
$\HS_{ab}$.  Our task here will be to remove that restriction and prove it for
a general density operator $\rho=\rho_{ab}$, where subscripts have been added
to avoid any confusion in the following argument.  To that end first
``purify'' $\rho_{ab}$ by introducing an auxiliary system $\HS_c$ and a ket
\begin{equation}
  \ket{\Phi} = \sum_q \sqrt{p_q}\ket{\psi_q}\ot\ket{c_q}
\label{Aeqn1}
\end{equation}
on $\HS_{abe}$, with $\{\ket{c_q}\}$ an orthonormal basis of $\HS_c$, and
coefficients $\{\ket{\psi_q}\}$ chosen so that
\begin{equation}
  \rho_{ab} = \Tr_c(\dya{\Phi}) = \sum_q p_q\dya{\psi_q}.
\label{Aeqn2}
\end{equation}

\Lem \emph{
Suppose that the $\VS=\{V_j\}$ information about $\ST_a$ is perfectly present
in $\ST_b$ for $\rho_{ab}$.  Then that is also true for every $\ket{\psi_q}$
in \eqref{Aeqn1} for which $p_q>0$. 
}

To show this, insert \eqref{Aeqn2} in place of $\rho$ on the right side of
\eqref{eqn6}, so each $\rho_{bj}$ is a sum over $q$ of positive operators
$\rho_{bqj}$.  Then in order that \eqref{eqn7} hold it is necessary (and
obviously sufficient) that $\rho_{bqj}\rho_{bq'k}=0$, for all $q$ and $q'$,
whenever $j\neq k$. Setting $q'=q$ gives the desired result.
Now apply the lemma to both the $\VS$ and the strongly incompatible $\WS$
type of information. Since for one of the pure states $\ket{\psi_q}$ on
$\HS_{ab}$ both types of information about $\ST_a$ are in $\ST_b$, theorem 4
of \cite{Grff05} tells us that for this pure state \emph{all} information
about $\ST_a$ is present in $\ST_b$.  This implies (and is implied by, see
theorem 3 (i) of \cite{Grff05}) that $\ket{\psi_q}$ is maximally entangled, or
\begin{equation}
  \Tr_b(\dya{\psi_q}) = I_a/d_a.
\label{Aeqn3}
\end{equation}

Next argue that the $\CS=\{\dya{c_q}\}$ type of information
about $\ST_c$ is absent from $\ST_a$ by showing that the conditional
density operators
\begin{equation}
  \Tr_c(\rho_{ac} \dya{c_q}) = p_q (I_a/d_a)
\label{Aeqn4}
\end{equation}
depend on $\ket{c_q}$ only through the numerical factor $p_q$; see
\eqref{eqn13} with systems appropriately renamed.  Indeed, \eqref{Aeqn4}
follows from \eqref{Aeqn1} when one replaces $\rho_{ac}$ with
$\Tr_b(\dya{\Phi})$ and interchanges the order of partial traces.  The same
argument applies if we use any other choice of orthonormal basis of $\HS_c$
for the expansion \eqref{Aeqn1}: as is well known, changing that basis does
not alter the density operator $\rho=\rho_{ab}$ we began with, but simply
expresses it in terms of a different ensemble.  Consequently, all types of
information about $\ST_c$ are absent from $\ST_a$, which is to say
$\rho_{ac}=\rho_a\ot\rho_c$, theorem 1 (iii) of \cite{Grff05}, and thus all
types of information about $\ST_a$ are also absent from $\ST_c$.

One more step is needed.  The presence of $\VS$ information about $\ST_a$ in
$\ST_b$ means it is also present in $\ST_{bc}$. (This is intuitively obvious,
but can also be shown formally from the definition in (35) of \cite{Grff05},
where one simply replaces $B^k$ with $B^k\ot I_c$.  Or one can use the
definition in \eqref{eqn7} of the present paper, along with the fact that if
$P$ and $Q$ are positive operators on $\HS_e\ot\HS_f$ with $PQ\neq 0$, then
$P_e Q_e\neq 0$, where $P_e$ and $Q_e$ are partial traces over $\HS_f$---use
the spectral representations and take traces.) Of course the same is true of
the $\WS$ information.  Hence, applying theorem 4 of \cite{Grff05} to the
bipartite system consisting of $\ST_a$ on the one hand and $\ST_{bc}$ on the
other, with $\ket{\Phi}$ an entangled ket on $\HS_a\ot\HS_{bc}$, we conclude,
from the presence of two strongly incompatible types of information, that all
the information about $\ST_a$ is present in the combined system $\ST_{bc}$.
From this and the absence of all information about $\ST_a$ from $\ST_c$
demonstrated earlier, it follows from theorem 8 (iii) of \cite{Grff05}, using
$\ket{\Phi}$ as a pure state pre-probability on $\HS_{abc}$, that all
information about $\ST_a$ is in $\ST_b$.  This completes the proof of the
theorem.

Next we will show that two orthonormal bases $\VS=\{\ket{v_j}\}$ and
$\WS=\{\ket{w_k}\}$ are strongly incompatible if and only if the graph
described in Sec.~\ref{sct5a} is connected.
First, assume $\VS$ and $\WS$ are \emph{not} strongly incompatible, so there
is a projector $P$ that commutes with every $\dya{v_j}$ and every $\dya{w_k}$,
and is neither 0 nor $I$, so projects onto a proper subspace $\PS$ of the
Hilbert space.  Then some subset of the $\{\ket{v_j}\}$ are inside $\PS$ and
the rest are in its orthogonal complement $\PS^\perp$, for otherwise $\PS$
would not commute with them, and the same is true of the $\{\ket{w_k}\}$.
Evidently there can be no nonzero $\inpd{v_j}{w_k}$ for a $\ket{v_j}$ (or
$\ket{w_k}$) in $\PS$ and a $\ket{w_k}$ (or $\ket{v_j}$) in $\PS^\perp$.
Consequently the graph cannot be connected, as it has at least two components,
one for $\PS$ and one for $\PS^\perp$ vertices.

For the converse, assume the graph is not connected, and renumber the vertices
so that the $\{\ket{v_j}\}$ with $1\leq j\leq m < d_a$ are all the $v$ vertices
in one connected component $\CS$ of the graph, with the others in its
complement.  The projector
\begin{equation}
  P=\sum_{j=1}^m \dya{v_j}
\label{Aeqn5}
\end{equation}
obviously commutes with all the $\{\ket{v_j}\}$, and in addition commutes with
every $\dya{w_k}$ when $\ket{w_k}$ is \emph{not} in $\CS$, as otherwise there
would be an edge from some $\ket{v_j}$ with $j\leq m$ to this $\ket{w_k}$.
Now apply the same argument with $I-P$ in place of $P$ to show that $I-P$
commutes with every $\dya{w_k}$ when $\ket{w_k}$ \emph{is} in $\CS$.  Since
$P$ and $I-P$ commute with the same things,  we have shown that $P$
commutes with all the $\dya{w_k}$ as well as the $\dya{v_j}$. As $P$
is neither 0 nor $I$, $\VS$ and $\WS$ are not strongly incompatible.

\section{Truncation Theorem and Related}
\label{sctpb}

If the $\VS=\{V_j\}$ information about $\ST_a$ is perfectly present in
$\ST_b$, there is, as noted following \eqref{eqn7}, a decomposition $\{T_k\}$
of $I_b$ such that
\begin{equation}
  \lgl V_j T_k\rgl = \dl_{jk} \lgl V_j\rgl.
\label{Beqn1}
\end{equation}
(Note that $V_j T_k$ is $V_j\ot T_k\ot I_c$.)
The equivalence is shown in
\cite{Grff05}, where in fact \eqref{Beqn1} is the primary definition.
If the pre-probability defining $\lgl\rgl$ is a pure state
$\ket{\Psi}\in\HS_{abc}$, \eqref{Beqn1} implies that
\begin{equation}
  V_j T_k\ket{\Psi} = 0 \text{ for } j\neq k.
\label{Beqn2}
\end{equation}
As a consequence, and using the fact that $I_a=\sum_j V_j$, $I_b=\sum_k T_k$,
we have
\begin{equation}
  V_j\ket{\Psi} = V_j T_j\ket{\Psi} = T_j\ket{\Psi},
\label{Beqn3}
\end{equation}
and
\begin{equation}
  \ket{\Psi} = \sum_j V_j T_j \ket{\Psi} = \sum_j V_j \ket{\Psi}.
\label{Beqn4}
\end{equation}
Therefore it follows that
\begin{equation}
 \lgl AC \rgl = \mte{\Psi}{\Bigl[\sum_{jk}V_j T_j AC V_k T_k\Bigr]}
 = \mte{\Psi}{\Bigl[\sum_j V_j  A V_j C\Bigr]},
\label{Beqn5}
\end{equation}
which is \eqref{eqn9} when the pre-probability is a pure state.  In
\eqref{Beqn5} we have used the fact that $T_j$ commutes with $AC$, $T_j T_k =
\dl_{jk} T_j$, and \eqref{Beqn4}.

To extend the argument to a general density operator $\rho$ on $\HS_{abc}$,
introduce a fictitious system $\HS_r$, purify $\rho$ to a ket
$\ket{\Psi}\in\HS_{abcr}$, and apply \eqref{Beqn5} to the three part system
consisting of $\ST_a$, $\ST_b$, and in place of $\ST_c$ the combined system
$\ST_{cr}$. The significance of $V_j$, $T_j$, and $A$ is the same as before,
while $C$ can be replaced by any operator on $\HS_{cr}$.  If in particular we
use $C\ot I_r$, the result is \eqref{eqn9}.  The equivalence of \eqref{eqn8}
and \eqref{eqn9} is a straightforward exercise when one notes that $\lgl
AC\rgl=\Tr(AC\rho_{ac})$, and that \eqref{eqn9} holds for all $A$ and $C$
(operating on $\HS_a$ and $\HS_c$).  This completes the proof.

The following example shows that the Exclusion theorem does not possess a
simple converse of the type mentioned in Sec.~\ref{sct5b}.  The entangled
state
\begin{equation}
  2\ket{\Psi} = \ket{000} + \ket{011} + \ket{100} - \ket{111}
\label{Beqn6}
\end{equation}
on $\HS_{abc}$, with qubits in the order $\ket{abc}$, has the property that
all information about $\ST_a$ is present in the combined system $\ST_{bc}$,
the $\XS$ information about $\ST_a$ is perfectly present in both $\ST_b$ and
in $\ST_c$, whereas both the $\YS$ (basis $\{(\ket{0}\pm i\ket{1})/\st\}$) and
the mutually unbiased $\ZS$ information about $\ST_a$ are perfectly absent
from both $\ST_b$ and from $\ST_c$.  Perhaps the easiest way to check this is
to expand $\ket{\Psi}$ in the $\XS$, $\YS$, and $\ZS$ bases of $\HS_a$ in
turn, and look at the coefficients in $\HS_{bc}$.  That all the information
about $\ST_a$ is in $\ST_{bc}$ follows from the observation that any one of
these expansions (and therefore all three) is in Schmidt form with Schmidt
coefficients of equal magnitude, so theorem 3(i) of \cite{Grff05} applies.
Note that we have an example in which if $\ST_a$ and $\ST_b$ are considered
two parts of a bipartite system with pre-probability given by the density
operator $\Tr_c(\dya{\Psi})$, one would be mistaken to suppose that the
absence of the two mutually unbiased types of information $\YS$ and $\ZS$
about $\ST_a$ from $\ST_b$ implied the complete absence of all information.
This confirms the remarks at the end of Sec.~\ref{sct5c}.

\section{Absence Theorems}
\label{sctpc}

For part (i), write $\ket{\Psi}$ in the form
\begin{equation}
  \ket{\Psi} = \sum_j \ket{v_j}\ot\ket{\bt_j},
\label{Ceqn1}
\end{equation}
where the $\ket{\bt_j}$ are expansion coefficients. The fact that the
$\{\ket{v_j}\}$ or $\{\dya{v_j}\}$ information about $\ST_a$ is absent from
$\ST_b$ means that the $\ket{\bt_j}$ are all proportional to one another, thus
multiples of $\ket{\bt_1}$, assuming it is nonzero.  Inserting
$\ket{\bt_j}=c_j\ket{\bt_1}$ in \eqref{Ceqn1} shows that
$\ket{\Psi}=\ket{a}\ot\ket{\bt_1}$ is a product state, so no information about
$\ST_a$ is in $\ST_b$ or vice versa.

To prove (ii) we employ an orthonormal basis $\{Q_r\}$, $0\leq r\leq d^2_a-1$
of the space $\hat\HS_a$ of linear operators on $\HS_a$, in the sense that
\begin{equation}
  \lgl Q_r,Q_s\rgl := (1/d_a)\Tr_a(Q_r\ad Q_s)=\dl_{rs},
\label{Ceqn2}
\end{equation}
with $Q_0=I_a$, and thus $\Tr_a(Q_r)=0$ for $r>0$. Expand $\rho$ as
\begin{equation}
  \rho = (1/d_a) \sum_r Q_r\ot B_r,
\label{Ceqn3}
\end{equation}
where, see \eqref{Ceqn2}, the expansion coefficients are given by
\begin{equation}
  B_r = \Tr_a(Q_r\rho),
\label{Ceqn4}
\end{equation}
with
\begin{equation}
  B_0 = \Tr_a(\rho) =  \rho_b.
\label{Ceqn5}
\end{equation}
The absence from $\ST_b$ of each species $V^{(m)}=\{V^{(m)}_j\}$ of
information about $\ST_a$ means that, see \eqref{eqn13}, 
\begin{equation}
  \Tr_a(V^{(m)}_j\rho) = p_{jm} \rho_b = p_{jm} B_0,
\label{Ceqn6}
\end{equation}
where the $p_{jm}$ are nonnegative constants.  By hypothesis, the collection
$\{V^{(m)}_j\}$ for all $m$ and all $j$ spans the operator space $\hat\HS_a$,
so any $Q_r$ can be written as a sum
\begin{equation}
  Q_r = \sum_{jm} c_{rjm} V^{(m)}_j,
\label{Ceqn7}
\end{equation}
with suitable coefficients $c_{rjm}$. (These may not be unique, but that does
not matter.) Insert \eqref{Ceqn7} in \eqref{Ceqn4} and use \eqref{Ceqn6} to
conclude that every $B_r$ is a multiple of $B_0$, and therefore
$\rho=\rho_a\ot B_0$ is a product.

The need for at least $d_a+1$ orthonormal bases of $\HS_a$ in order to check
that all information about $\ST_a$ is absent from $\ST_b$ can be seen in the
following way.  Each basis of $\HS_a$ gives rise to $d_a$ orthogonal, and
hence linearly independent, operators in the $d^2_a$-dimensional space
$\hat\HS_a$.  But these $d_a$ projectors sum to the identity $I$ for each such
basis, and therefore $\nu$ such bases will give rise to at most $\nu(d_a-1)+1$
linearly independent operators, which is $d^2_a$ when $\nu=d_a+1$.

\section{Generalized No Cloning}
\label{sctpd}

From \eqref{eqn16} one sees that for every $j$ and $k$ in $J$,
\begin{equation}
  |\inpd{\gm_j}{\gm_k}|\cdot |\inpd{\bt_j}{\bt_k}| 
  = |\mted{\al_j}{M\ad M}{\al_k}| = |\inpd{\al_j}{\al_k}|,
\label{Deqn1}
\end{equation}
and therefore, in view of \eqref{eqn17},
\begin{equation}
  |\inpd{\gm_j}{\gm_k}| = 1
\label{Deqn2}
\end{equation}
whenever $j\neq k$ and $|\inpd{\al_j}{\al_k}|\neq 0$.  The pairs $(j,k)$ for
which \eqref{Deqn2} holds form, by hypothesis, a connected graph on $J$, which
means that the normalized kets $\ket{\gm_j}$ are identical apart from phase
factors, so each is a multiple of just one of them; let us say
$\ket{\gm_j}=e^{i\phi_j}\ket{\gm_1}$. Replace the right side of \eqref{eqn16}
with $\ket{\bt'_j}\ot\ket{\gm_1}$, where $\ket{\bt'_j}=e^{i\phi_j}
\ket{\bt_j}$, and define the linear operator $U$ on $\GS_a$ so that
\begin{equation}
  U\ket{\al_j} = \ket{\bt'_j}.
\label{Deqn3}
\end{equation}
Since $|\inpd{\bt'_j}{\bt'_k}|=|\inpd{\bt_j}{\bt_k}|=|\inpd{\al_j}{\al_k}|$,
$U$ is unitary as a map from $\GS_a$ to $\GS_b$.


\begin{thebibliography}{10}
\bibitem{ntk00}
A number of paradoxes are discussed in Chs.~19 to 25 of \cite{Grff02c}.
\bibitem{ntk05}
The claim that decoherence by itself solves the measurement problem has, in our
  opinion, been overstated; see the clear critique in \cite{Adlr03}. Instead,
  the usual treatments of decoherence suffer from the same internal
  inconsistency that constitutes the infamous ``measurement'' problem of
  quantum foundations, as discussed by Wigner \cite{Wgnr63} and Mittelstaedt
  \cite{Mttl98} among others. Solving the measurement problem requires that one
  introduce probabilities in quantum theory in a consistent manner, as
  discussed in \cite{Grff02c}, and then decoherence can be consistently
  discussed as a quantum mechanical process: see
  \cite{GMHr90,GMHr93,Omns97,Omns99,Omns05} and Ch.~26 of \cite{Grff02c}.
\end{thebibliography}

\end{document}